\documentclass[journal,10pt]{IEEEtran}

\usepackage{amsmath,amssymb,amsfonts}
\usepackage{graphicx}
\usepackage{tikz}
\usetikzlibrary{arrows.meta,positioning,decorations.pathreplacing,calc}
\usepackage{cite}
\usepackage{url}
\usepackage{xcolor}
\usepackage{booktabs}

\newcommand{\E}{\mathbb{E}}

\newcommand{\jmax}{g_{\mathrm{FAS}}}

\begin{document}

\title{Fluid Antenna-Aided Noise Modulation:\\ Spatial Diversity for Variance-Based Wireless Communication}

\author{Hadi Zayyani, Felipe A. P. de Figueiredo, Pedro M. R. Pereira, Fernando D. A. García, Rausley A. A. de Souza,
\thanks{Authors are with INATEL, Brazil.}}

\maketitle

\begin{abstract}
Noise modulation (NoiseMod) encodes information in the \emph{variance} of a
transmitted noise-like waveform rather than in its amplitude, phase, or
frequency, and is attractive for ultra-low-power and covert links. Its main
weakness is that, unlike classical modulation, it exhibits \emph{no}
diversity under Rayleigh fading: its bit error probability (BEP) decays only
as $1/(N_s\delta)$, where $N_s$ is the number of noise samples per bit and
$\delta$ the useful-to-thermal noise variance ratio. Independently, fluid
antenna systems (FAS) have been shown to recover substantial selection
diversity from a single radiating element that switches among $N_p$ closely
spaced ports, without extra radio-frequency chains. This paper combines the
two: we equip a NoiseMod receiver with a fluid antenna and derive its
average BEP. For idealized, mutually independent ports we obtain an exact
closed-form BEP via order statistics of the port envelopes. For the
physically accurate, spatially correlated case governed by Jake's model, we
build a semi-analytical BEP using the two-stage channel approximation of
Khammassi \emph{et al.} We validate both regimes against full signal-level
Monte Carlo simulation and show that (i) FAS restores a diversity order that
grows with the number of ports $N_p$ when ports are weakly correlated, (ii)
this gain saturates once the fluid-antenna aperture $W\lambda$ is fixed and
$N_p$ grows, mirroring the outage-probability saturation reported for FAS,
now observed for BEP, and (iii) an intrinsic, correlation-independent (and
$\delta$-independent) BEP floor set only by $N_s$ and the variance ratio
$\alpha$ persists regardless of the antenna diversity order.
\end{abstract}


\section{Introduction}

Since the earliest days of digital radio, information has been carried in
the amplitude, phase, or frequency of a deterministic carrier, with noise
treated purely as an impairment to be suppressed. Noise modulation
(NoiseMod) inverts this premise: information is embedded in the
\emph{statistics} -- specifically the variance -- of a transmitted
noise-like waveform, so that a receiver decides the bit by comparing an
estimated sample variance against a threshold \cite{basar_noisemod}. The
concept traces back to Kish's zero-power stealth-communication idea using
switched resistor noise \cite{kish2005} and was placed on a rigorous
communication-theoretic footing, together with a thermal-noise
implementation (TherMod), in \cite{basar_thermod}. Because NoiseMod needs no
stable oscillator or power amplifier, it is attractive for ambient/backscatter
IoT sensors and covert links, and has since been extended to on-off and
ternary signaling \cite{onoff_noisemod, ternary_noisemod} and surveyed in
\cite{survey_noisebased}.

A central finding of \cite{basar_noisemod} is that NoiseMod, unlike
amplitude/phase modulation, extracts \emph{no diversity} from a single
Rayleigh-fading branch: its average BEP decays only linearly with the
inverse of the useful-noise-to-thermal-noise ratio $\delta$, i.e.\
$\bar P_b\propto 1/(N_s\delta)$, in stark contrast with the exponential decay
obtained under a pure AWGN channel. The letter partially remedies this with a
\emph{time-diversity} variant (TD-NoiseMod) that spreads the $N_s$ samples of
one bit over $I$ independent fading blocks, at the cost of $I\times$ longer
bit duration (i.e., a proportional loss in information rate and increased
latency).

Independently, fluid antenna systems (FAS) have emerged as a way to harvest
spatial diversity from a \emph{single} radiating element that can switch
instantaneously among $N_p$ ports distributed over a line of length
$W\lambda$ \cite{wong2021fas}, avoiding the extra RF chains that classical
multi-antenna selection combining requires. The first-order statistics of
FAS were characterized in \cite{wong2021fas}, later refined with tighter
outage and diversity-gain results in \cite{new2024insights}, and the
underlying spatial correlation -- originally simplified -- was shown by
Khammassi \emph{et al.} \cite{khammassi2023} to depart from the
physically accurate Jake's model \cite{stuber2018}; they proposed a
two-stage approximation that restores Jake's correlation while keeping the
port-selection CDF analytically tractable via a single radial integral. A
further block-correlation refinement appears in \cite{ramirez2024block}, and
practical limited-observation port-selection strategies are studied in
\cite{chai2022port}.

FAS spends a \emph{spatial} resource (an antenna aperture of a few
wavelengths) to obtain diversity, whereas TD-NoiseMod spends a
\emph{temporal} one. Since a NoiseMod transmitter/receiver pair already
values low complexity and low latency, replacing or augmenting temporal
repetition with spatial port selection is a natural fit: no extra bit
duration is required, and the fluid antenna control mechanism is
independent of the modulation format. Somewhat surprisingly, this
combination does not appear to have been analyzed previously (FAS has been
combined with index modulation \cite{zhu2024fas_im}, but not with
variance-based signaling). This paper closes that gap. Our contributions
are:
\begin{itemize}
\item We formulate FAS-aided NoiseMod, in which the receiver performs ideal
selection combining of the port with the largest instantaneous envelope
before variance-based bit detection, and derive its conditional and average
BEP.
\item For mutually independent ports (the idealized large-aperture limit) we
derive an \emph{exact} closed-form envelope distribution via order
statistics, giving a numerically exact average BEP that we validate against
Monte Carlo simulation to within simulation accuracy.
\item For ports correlated according to Jake's model we build a
semi-analytical BEP from the two-stage approximation of
\cite{khammassi2023}. Comparing it against Monte Carlo simulation, we
identify and report a regime -- moderate port count with weak correlation --
in which the closed-form optimal repetition parameter $R^\star$ of
\cite{khammassi2023} collapses to unity and the tool loses sensitivity to
the aperture $W$; this had not been previously exposed because outage
probability (the metric of \cite{khammassi2023}) is a bulk-CDF quantity,
whereas averaging a Q-function-shaped BEP is dominated by the extreme lower
tail of the envelope distribution.
\item We show, via extensive Monte Carlo simulation, that the achievable BEP
\emph{saturates} as $N_p$ grows for fixed aperture $W$ -- the BEP analogue of
the outage-probability saturation of \cite{khammassi2023} -- and we identify
an intrinsic, $N_p$-independent (and $\delta$-independent) BEP floor set
purely by $N_s$ and $\alpha$.
\end{itemize}

\section{Preliminaries}

\subsection{Noise Modulation}
\label{ssec:noisemod}
In NoiseMod \cite{basar_noisemod}, a bit is conveyed by transmitting $N_s$
samples of a zero-mean circularly-symmetric Gaussian waveform with variance
$\sigma_0^2$ (bit-0) or $\sigma_1^2=\alpha\sigma_0^2$ (bit-1), $\alpha>1$.
Over a channel with (possibly fading) coefficient $h$ and AWGN of variance
$\sigma_w^2$, the $n$-th received baseband sample is
\begin{equation}
s_n = h\,r_n + w_n,\qquad n=1,\dots,N_s,
\label{eq:sn}
\end{equation}
with $r_n\!\sim\!\mathcal{CN}(0,\sigma_b^2)$, $b\in\{0,1\}$, and
$w_n\!\sim\!\mathcal{CN}(0,\sigma_w^2)$. Defining $\delta=\sigma_0^2/\sigma_w^2$,
the receiver estimates $\hat\sigma_s^2=\frac1{N_s}\sum_n|s_n|^2$ and compares
it to a threshold $\gamma=\eta\sigma_w^2$. For known $|h|^2$, choosing $\eta$
to equalize the two conditional error probabilities gives \cite{basar_noisemod}
\begin{equation}
\eta = \frac{2CD}{C+D},\quad C=1+|h|^2\delta,\ \ D=1+|h|^2\alpha\delta,
\label{eq:eta}
\end{equation}
and the resulting conditional BEP is
\begin{equation}
P_b\!\left(|h|^2\right)= Q\!\left(\frac{\sqrt{N_s}\,|h|^2\delta(\alpha-1)}
{2+|h|^2\delta(\alpha+1)}\right).
\label{eq:condBEP}
\end{equation}
Under Rayleigh fading, $|h|^2$ is exponentially distributed and averaging
\eqref{eq:condBEP} yields a BEP that decays only as $1/(N_s\delta)$
\cite{basar_noisemod} -- a symptom of the fact that \eqref{eq:condBEP} does
not vanish as $\delta\to\infty$ for a \emph{fixed} $|h|^2$ close to its
typical (small) Rayleigh value, and the exponential tail of $|h|^2$ offers
no protection.

\subsection{Fluid Antenna Channel Model}
\label{ssec:fas}
An FAS receiver realizes $N_p$ ports equally spaced over a line of length
$W\lambda$; the channel gains
$\mathbf g=(g_1,\dots,g_{N_p})^T\sim\mathcal{CN}(\mathbf 0,\boldsymbol\Sigma_g)$
follow Jake's model,
$(\boldsymbol\Sigma_g)_{k,\ell}=\sigma^2 J_0\!\big(2\pi(k-\ell)W/(N_p-1)\big)$
\cite{stuber2018}. With ideal (instantaneous) port switching, the receiver
observes
\begin{equation}
\jmax = \max\{|g_1|,\dots,|g_{N_p}|\}.
\label{eq:gfas}
\end{equation}
Because \eqref{eq:gfas} mixes $N_p$ correlated Rician-type conditional
distributions, its exact CDF requires $N_p$-fold integration for $N_p>3$
\cite{zhang2002}. Khammassi \emph{et al.} \cite{khammassi2023} approximate
$\mathbf g$ in two stages: (i) a low-rank-plus-residual representation keeps
only the $\varepsilon\text{-rank}<N_p$ dominant eigenpairs
$(s_l,\mathbf u_l)$ of $\boldsymbol\Sigma_g$ exceeding a threshold
$\varepsilon$, reducing the CDF to $2\varepsilon\text{-rank}$ nested
integrals; (ii) a second approximation replaces an $N_p$-column random
matrix by one with independent rows, yielding a genuinely single-integral
CDF
\begin{align}
F_{\Psi_R}(g)=\!\prod_{k=1}^{N_p}\!\int_0^{\infty}\!\!\frac{e^{-r/w_k}}{w_k}
\Big[1-Q_1\!\Big(\sqrt{\tfrac{2r}{\rho_k}},\sqrt{\tfrac{2}{\rho_k}}\,g\Big)\Big]^{R}\!dr,
\label{eq:FPsiR}
\end{align}
with $w_k=\sum_{l\le\varepsilon\text{-rank}}s_l u_{k,l}^2$,
$\rho_k=\sigma^2-w_k$, $Q_1(\cdot,\cdot)$ the first-order Marcum-Q function,
and $R$ a repetition order. The FAS envelope CDF is then approximated as
$F_{\jmax}(g)\approx\big(F_{\Psi_R}(g)\big)^{1/R}$, with the closed-form
choice $R^\star=\min\!\big(\big\lfloor 1.52(N_p-1)/(2\pi W)\big\rfloor,
N_p\big)$ shown to minimize a covariance-mismatch bound
\cite{khammassi2023}. We use \eqref{eq:FPsiR} directly in
Section~\ref{ssec:corrbep}.

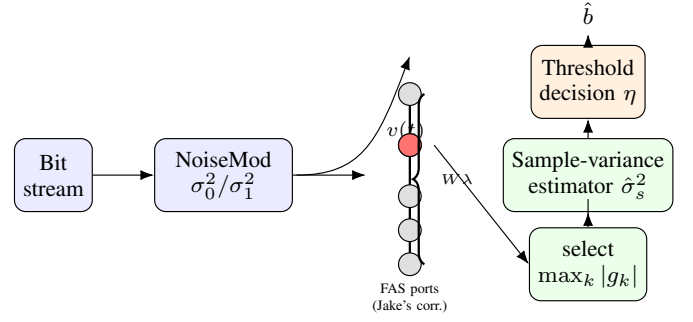
\begin{figure}[t]
\centering
\resizebox{\columnwidth}{!}{
\begin{tikzpicture}[
  font=\footnotesize,
  block/.style={draw, rounded corners, minimum height=0.9cm, align=center, inner sep=3pt},
  port/.style={draw, circle, minimum size=0.28cm, inner sep=0pt},
]
\node[block, fill=blue!8] (bits) at (0,0) {Bit\\stream};
\node[block, fill=blue!8, minimum width=1.7cm] (nm) at (2.1,0) {NoiseMod\\$\sigma_0^2/\sigma_1^2$};
\node[align=center] (wave) at (4.35,0.55) {\scriptsize $v(t)$};
\draw[-{Latex}] (bits) -- (nm);
\draw[-{Latex}] (nm.east) -- ++(0.9,0) node[midway,above,font=\tiny]{};

\begin{scope}[shift={(4.4,-1.1)}]
\draw[thick] (0,0) -- (0,2.1);
\foreach \y/\lab in {0/1,0.42/2,0.84/{$\cdots$},1.47/{$k^\ast$},2.1/{$N_p$}}{
  \node[port, fill=gray!25] at (0,\y) {};
}
\node[port, fill=red!55] at (0,1.47) {};
\draw[decorate,decoration={brace,amplitude=4pt},thick] (0.18,0) -- (0.18,2.1)
  node[midway,right=2pt,font=\tiny]{$W\lambda$};
\node[font=\tiny, align=center] at (0,-0.42) {FAS ports\\(Jake's corr.)};
\end{scope}
\draw[-{Latex}] (3.0,0) .. controls (3.8,0) and (4.0,0.4) .. (4.4,1.47*1 - 1.1 + 1.1);
\node[block, fill=green!8, align=center] (sel) at (6.6,-1.1) {select\\$\max_k|g_k|$};
\draw[-{Latex}] (4.7,-1.1+1.47) -- (sel.west);

\node[block, fill=green!8, minimum width=1.9cm] (bb) at (6.6,0) {Sample-variance\\estimator $\hat\sigma_s^2$};
\draw[-{Latex}] (sel) |- (bb.south);

\node[block, fill=orange!12] (dec) at (6.6,1.15) {Threshold\\decision $\eta$};
\draw[-{Latex}] (bb) -- (dec);
\node (out) at (6.6,2.0) {$\hat b$};
\draw[-{Latex}] (dec) -- (out);
\end{tikzpicture}}
\caption{FAS-aided NoiseMod: the transmitter is unmodified; the receiver
fluid antenna selects the port of largest instantaneous envelope before
variance-based threshold detection.}
\label{fig:system}
\end{figure}

\section{System Model: FAS-Aided NoiseMod}
\label{sec:system}

Fig.~\ref{fig:system} illustrates the proposed architecture. The
NoiseMod transmitter is unchanged. At the receiver, an FAS with $N_p$ ports
spans $W\lambda$; the port channels follow Jake's model as in
Section~\ref{ssec:fas}. We adopt the standard idealization used throughout
the FAS literature \cite{wong2021fas,khammassi2023,new2024insights} of an
ideal, delay-free switch to the port with the largest instantaneous
envelope -- practical limited-observation port selection strategies exist
\cite{chai2022port} and are compatible with, but not the focus of, our
analysis (see Section~\ref{ssec:discussion}). Since $r_n$ and $w_n$ in
\eqref{eq:sn} are circularly symmetric, only the magnitude of the channel
matters for variance-based detection; we may therefore replace $h$ in
\eqref{eq:sn}--\eqref{eq:eta} with the real, non-negative selected envelope
\begin{equation}
h \;\longrightarrow\; \jmax=\max_{1\le k\le N_p}|g_k|,
\end{equation}
so that the FAS-aided conditional BEP follows directly from
\eqref{eq:condBEP}:
\begin{equation}
P_b(u) = Q\!\left(\frac{\sqrt{N_s}\,u\,\delta(\alpha-1)}{2+u\,\delta(\alpha+1)}\right),
\qquad u \triangleq \jmax^2 .
\label{eq:condBEP_fas}
\end{equation}
Equation~\eqref{eq:condBEP_fas} requires the receiver to know
$\jmax^2$ (equivalently, to estimate the selected port's channel gain) to
set the threshold via \eqref{eq:eta}; this is the same channel-state
requirement as coherent (threshold-adaptive) NoiseMod in
\cite{basar_noisemod}, now applied to a single scalar (the selected port),
not to all $N_p$ ports.

\section{Average BEP Analysis}
\label{sec:bepanalysis}
The average BEP is $\bar P_b=\E_U[P_b(U)]=\int_0^\infty P_b(u)\,dF_U(u)$,
where $F_U(u)=F_{\jmax}(\sqrt u)$. Because $P_b(u)$ has no simple moment
generating function counterpart here (it is not the standard Q-function of a
linear SNR, but of a saturating rational function of $u$), we evaluate the
Riemann--Stieltjes integral numerically once $F_{\jmax}$ is known; this
requires only a single outer numerical integration in every case below,
regardless of $N_p$.

\subsection{Independent-Port Bound}
\label{ssec:indepbep}
When the $N_p$ ports are mutually independent (the limit of a large
aperture $W$, or a benchmark for the best possible diversity at fixed
$N_p$), each $|g_k|^2$ is i.i.d.\ exponential with mean $\sigma^2$, so
$U=\jmax^2=\max_k|g_k|^2$ is the maximum of $N_p$ i.i.d.\ exponential
variables, with \emph{exact} closed-form CDF and density
\begin{align}
F_U(u) &= \big(1-e^{-u/\sigma^2}\big)^{N_p}, \label{eq:FUindep}\\
f_U(u) &= \frac{N_p}{\sigma^2}\,e^{-u/\sigma^2}\big(1-e^{-u/\sigma^2}\big)^{N_p-1}.
\label{eq:fUindep}
\end{align}
The average BEP follows as the single integral
$\bar P_b^{\,\mathrm{indep}}=\int_0^\infty P_b(u) f_U(u)\,du$, evaluated
numerically. Unlike the $N_p=1$ case, where $\bar P_b\propto 1/(N_s\delta)$
\cite{basar_noisemod}, order statistics push the effective operating point
$U$ towards larger values as $N_p$ grows, so $\bar P_b^{\,\mathrm{indep}}$
falls off much faster than $1/(N_s\delta)$, as confirmed numerically in
Section~\ref{sec:results}.

\subsection{Correlated Ports Under Jake's Model}
\label{ssec:corrbep}
For finite $W$, ports are correlated and \eqref{eq:FUindep} is optimistic.
We use the two-stage approximation of \cite{khammassi2023} recalled in
Section~\ref{ssec:fas}: for a given $(N_p,W)$ we compute
$\boldsymbol\Sigma_g$, its $\varepsilon\text{-rank}$ dominant eigenpairs
(threshold $\varepsilon=\sigma^2/2N_p$, as recommended in
\cite{khammassi2023}), the associated $R=R^\star$, and evaluate
$F_{\jmax}(g)\approx\big(F_{\Psi_R}(g)\big)^{1/R}$ from \eqref{eq:FPsiR} on a
grid of $g$; the average BEP is then obtained by integrating $F_U(u)=F_{\jmax}(\sqrt
u)$ against the (analytically known) derivative of $P_b(u)$, using
integration by parts,\footnote{Writing $h(u)=P_b(u)$, which is
non-increasing with limit $h(\infty)=P_b^\infty$ (Section~\ref{ssec:floor}),
$\bar P_b=\E[h(U)]=h(\infty)+\int_0^\infty(-h'(u))F_U(u)\,du$. We found this
form far better conditioned numerically than a direct Riemann--Stieltjes
sum on a fixed grid, which convergence testing showed to be
under-resolved unless the grid is prohibitively fine; adaptive quadrature
on the form above converges reliably.} rather than a fixed-grid sum.
Evaluating \eqref{eq:FPsiR} requires the first-order Marcum-Q function
$Q_1(a,b)$ at noncentrality $a^2$ that can reach $10^5$--$10^6$ when the
residual variance $\rho_k$ is small (i.e., when $\varepsilon\text{-rank}$
already captures nearly all of the port's variance); we evaluate this
exactly via the noncentral-$\chi^2$ series whenever it does not overflow
(validated against an independent, numerically-stable reference built from
the exponentially-scaled Bessel function to agree to machine precision up
to $a^2\sim10^8$) and fall back to the classical moment-matched Gaussian
approximation of the noncentral-$\chi^2$ tail only on the rare inputs
(numerically zero $b$ together with sizeable $a$) where the series itself
overflows.

\emph{A caveat on accuracy.} Section~\ref{sec:results} shows that this
semi-analytical BEP tracks Monte Carlo simulation closely only for the
most strongly-correlated apertures tested (small $W$, large $R^\star$); it
degrades as $W$ grows, first gradually (a factor of $\sim\!1.5$--$2.4$
under-estimate at $R^\star=2$) and then qualitatively once $R^\star$ rounds
down to $1$ at moderate $N_p$: the repetition-order mechanism that makes
\eqref{eq:FPsiR} single-integral tractable then adds no further averaging,
and the approximation \emph{loses sensitivity to $W$ itself}, collapsing to
the same curve for every aperture with $R^\star=1$ (a $2$--$4\times$
under-estimate at the well-sampled points we checked, though we would not
trust a tighter bound: other $(N_p,W)$ combinations we probed outside the
paper's main figures showed under-estimates of up to several hundred-fold
at $R^\star=1$ (e.g.\ $\sim\!330\times$ at $N_p=16,W=2$), so no single
multiplicative bound holds uniformly). This had not been visible in
\cite{khammassi2023}, whose validation targets outage probability (governed
by the bulk of $F_{\jmax}$) rather than an averaged Q-function BEP, which is
dominated by the extreme lower tail of $\jmax$ where the approximation is
weakest.

\subsection{Diversity Behavior and the Intrinsic NoiseMod Floor}
\label{ssec:floor}
As $u\to\infty$ (a perfect, noiseless channel to the selected port),
\eqref{eq:condBEP_fas} does \emph{not} vanish: it saturates at
\begin{equation}
P_b^{\infty} = \lim_{u\to\infty}P_b(u) = Q\!\left(\sqrt{N_s}\,\frac{\alpha-1}{\alpha+1}\right),
\label{eq:floor}
\end{equation}
independent of $\delta$. This is an intrinsic detection floor of
variance-based signaling with a finite sample count $N_s$: no amount of
spatial (or temporal) diversity can push $\bar P_b$ below $P_b^\infty$,
which can only be lowered by increasing $N_s$ (more observation
time/bandwidth per bit) or $\alpha$ (a larger variance contrast, at an
energy cost). For our operating parameters ($N_s=120,\alpha=10$),
$P_b^{\infty}=Q(8.96)\approx 10^{-19}$, far below the BEP range studied
here, so the saturation visible in Section~\ref{sec:results} is entirely
due to the FAS envelope distribution itself (Section~\ref{ssec:corrbep}),
not \eqref{eq:floor}; the two mechanisms are conceptually distinct and would
only compete at much larger $N_p$ or $W$.

\section{Numerical Results}
\label{sec:results}
Unless stated otherwise we use $N_s=120$, $\alpha=10$, and $\sigma^2=1$,
matching the parameters of \cite{basar_noisemod}. All semi-analytical curves
are cross-validated by a full signal-level Monte Carlo simulator that
generates correlated port channels via a Cholesky factorization of
$\boldsymbol\Sigma_g$, draws the $N_s$ noise-modulated samples of
\eqref{eq:sn} explicitly, forms $\hat\sigma_s^2$, and applies the threshold
of \eqref{eq:eta}; trial counts are chosen adaptively (from $2\times10^4$ to
$10^6$ bits) to keep at least tens of observed error events where feasible
(the lowest-BEP points on some curves rest on only 1--2 observed errors and
are flagged as such below).

\subsection{Independent Ports: Diversity Gain}
\begin{figure}[t]
\centering
\includegraphics[width=\columnwidth]{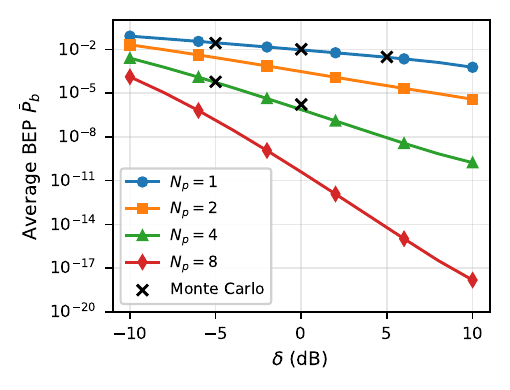}
\caption{Average BEP vs.\ $\delta$ for independent FAS ports (exact
closed form), $N_s=120$, $\alpha=10$. Markers: Monte Carlo (decorrelated
limit, $W=1000$).}
\label{fig:indep}
\end{figure}

Fig.~\ref{fig:indep} shows $\bar P_b^{\,\mathrm{indep}}$ versus $\delta$ for
$N_p\in\{1,2,4,8\}$, together with Monte Carlo points. The $N_p=1$ curve
reproduces the $1/(N_s\delta)$ trend of \cite{basar_noisemod}; already
$N_p=2$ visibly steepens the slope, and $N_p=8$ reduces the BEP by close to
$15$ orders of magnitude (a factor of $\sim\!4\times10^{14}$) at
$\delta=10$~dB relative to $N_p=1$. The
Monte Carlo markers match the closed-form curves to within simulation
noise across the observable range, confirming
\eqref{eq:FUindep}--\eqref{eq:fUindep}.

\subsection{Correlated Ports: Effect of the Aperture $W$}
\label{ssec:corrW}
\begin{figure}[t]
\centering
\includegraphics[width=\columnwidth]{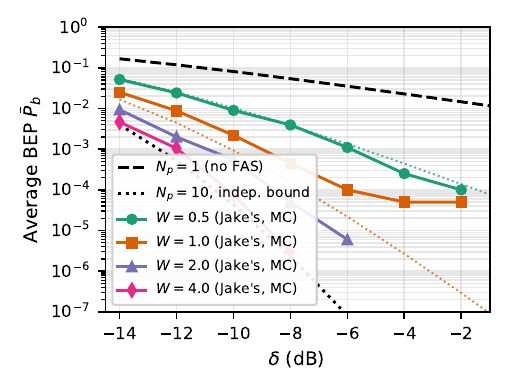}
\caption{Average BEP vs.\ $\delta$ for Jake's-correlated FAS ports,
$N_p=10$, $N_s=120$, $\alpha=10$. Solid: Monte Carlo. Thin dotted:
semi-analytical (Sec.~\ref{ssec:corrbep}), shown where $R^\star>1$.}
\label{fig:corr}
\end{figure}

Fig.~\ref{fig:corr} fixes $N_p=10$ and sweeps $\delta$ for
$W\in\{0.5,1,2,4\}$, obtained by Monte Carlo (solid lines/markers), bracketed
by the $N_p=1$ baseline (dashed) and the $N_p=10$ independent bound
(dotted). All correlated curves lie strictly between the two references, as
expected: correlation always costs diversity relative to the independent
bound, and reducing $W$ pulls the curve towards the no-FAS baseline. The
semi-analytical curves of Section~\ref{ssec:corrbep} (thin dotted, shown for
$W\in\{0.5,1\}$ where $R^\star\in\{4,2\}$) track the simulation closely for
$W=0.5$ (within $\sim\!20\%$ over the well-sampled portion of the curve,
tens or more observed error events) but noticeably under-estimate it for
$W=1$ (by a factor of $\sim\!1.5$--$2.4$ over the same well-sampled range);
the lowest one or two BEP points on each Monte Carlo curve rest on only
1--2 observed errors and are themselves too noisy to judge the
approximation against. Accuracy degrades with $R^\star$ but not simply
monotonically with it: it is tight at $R^\star=4$ ($W=0.5$) and visibly
looser at $R^\star=2$ ($W=1$); at $R^\star=1$ ($W\in\{2,4\}$, omitted from the
plot) the approximation collapses to the \emph{same} curve for both
apertures -- the degeneracy noted in Section~\ref{ssec:corrbep} -- which at
the well-sampled points under-estimates the true simulated BEP by roughly
$2$--$4\times$ for $W=2$ and by a comparable-to-smaller margin for $W=4$
(where the shared, degenerate curve happens to sit closer to the true,
lower $W=4$ curve). The qualitative lesson -- $R^\star$ collapsing to $1$
destroys the tool's ability to distinguish different apertures -- matters
more here than any single multiplicative error factor, consistent with the
caveat of Section~\ref{ssec:corrbep}.

\subsection{Saturation with the Number of Ports}
\begin{figure}[t]
\centering
\includegraphics[width=\columnwidth]{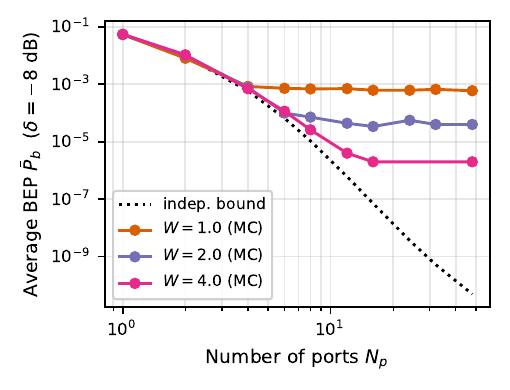}
\caption{Average BEP vs.\ number of ports $N_p$ at fixed $\delta=-8$~dB
(Monte Carlo), showing correlation-induced saturation for fixed aperture
$W$.}
\label{fig:sat}
\end{figure}

Fig.~\ref{fig:sat} is the central result of this study: it fixes
$\delta=-8$~dB and sweeps $N_p$ from 1 to 48 for $W\in\{1,2,4\}$ (Monte
Carlo), together with the independent bound. For every fixed $W$, the BEP
initially falls quickly with $N_p$ -- exactly the diversity gain motivating
FAS -- but then \emph{saturates}: at $W=1$ the floor is reached by
$N_p\approx6$ near $\bar P_b\approx7\times10^{-4}$, already an order of
magnitude above the independent bound at that same $N_p$, a gap that
widens to more than seven orders of magnitude by $N_p=48$ as the independent
bound keeps falling while the correlated curve stays flat; increasing the
aperture to $W=2$ and $W=4$ lowers the floor (to $\approx4\times10^{-5}$ and
a few$\times10^{-6}$, respectively -- the latter at the edge of what
$5\times10^5$ trials can resolve) and delays the point of saturation,
but never removes it while $W$ is fixed. This is the BEP counterpart of the
outage-probability saturation reported for FAS in \cite{khammassi2023}: once
the port spacing $W\lambda/(N_p-1)$ falls below roughly half a wavelength,
additional ports are so strongly correlated with existing ones that they
contribute negligible new information about the channel, and the
achievable diversity order is governed by $W$ (equivalently, by the number
of dominant eigenvalues $\varepsilon\text{-rank}\approx\lceil
3.19\,W N_p/(N_p-1)\rceil$ of $\boldsymbol\Sigma_g$ \cite{khammassi2023})
rather than by the raw port count $N_p$.

\subsection{Practical Discussion}
\label{ssec:discussion}
Two idealizations deserve comment. First, ideal port selection assumes the
receiver can instantaneously identify the strongest port; in practice this
requires either fast electronic switching with full-port SNR observation, or
a limited-observation strategy such as \cite{chai2022port}, which reports
that observing only $10\%$ of the ports already yields most of the outage
reduction -- a promising avenue to reduce the channel-sounding overhead of
FAS-NoiseMod as well, left for future work. Second, the coherent threshold
\eqref{eq:eta} requires estimating $\jmax^2$; because only one (scalar)
channel needs to be tracked -- the selected port's gain, not all $N_p$
gains simultaneously -- this is no more demanding than the channel
estimation already assumed for coherent (non-NC) NoiseMod
\cite{basar_noisemod}. Finally, spatial (FAS) and temporal (TD-NoiseMod)
diversity are complementary: FAS adds no latency or rate penalty but
saturates with the physical aperture, whereas TD-NoiseMod scales with the
time-spreading factor $I$ but multiplies the bit duration by $I$; combining
both is a natural extension when neither resource alone suffices.

\section{Conclusion}
We introduced FAS-aided noise modulation, combining ideal fluid-antenna port
selection with variance-based bit detection, and derived its average BEP
both for the idealized independent-port limit (exact, closed form) and for
the physically accurate Jake's-correlated case (semi-analytical, built on
the two-stage FAS channel approximation of Khammassi \emph{et al.}). Full
signal-level Monte Carlo simulation confirms that FAS restores meaningful
diversity to a scheme that otherwise has none, but that this gain saturates
once the antenna aperture $W\lambda$ is fixed, mirroring the outage
saturation known for FAS and now demonstrated for BEP; an additional,
$N_p$-independent floor set by the number of noise samples per bit remains
regardless of antenna diversity. These results suggest FAS as a
low-complexity, latency-free complement to temporal diversity for
low-power NoiseMod receivers, and motivate future work on joint FAS port
selection with limited channel observation, and on combined
spatio-temporal (FAS + TD-NoiseMod) designs.


\begin{thebibliography}{99}

\bibitem{basar_noisemod}
E. Basar, ``Noise modulation,'' \emph{IEEE Wireless Commun. Lett.}, vol. 13,
no. 3, pp. 844--848, Mar. 2024.

\bibitem{basar_thermod}
E. Basar, ``Communication by means of thermal noise: Towards networks with
extremely low power consumption,'' \emph{IEEE Trans. Commun.}, vol. 71, no.
2, pp. 688--699, Feb. 2023.

\bibitem{kish2005}
L. B. Kish, ``Stealth communication: Zero-power classical communication,
zero-quantum quantum communication and environmental-noise communication,''
\emph{Appl. Phys. Lett.}, vol. 87, no. 23, p. 234109, Dec. 2005.

\bibitem{mucchi2022}
L. Mucchi \emph{et al.}, ``Security and reliability performance of
noise-loop modulation: Theoretical analysis and experimentation,'' \emph{IEEE
Trans. Veh. Technol.}, vol. 71, no. 6, pp. 6335--6350, Jun. 2022.

\bibitem{onoff_noisemod}
A. A. dos Anjos and H. S. Silva, ``On-off digital noise modulation,''
\emph{IEEE Wireless Commun. Lett.}, vol. 14, no. 11, pp. 3595--3599, Nov.
2025.

\bibitem{ternary_noisemod}
M. K. Alshawaqfeh, O. S. Badarneh, Y. H. Al-Badarneh, M. T. Dabiri, and M. O.
Hasna, ``Thermal noise modulation: Optimal detection and performance
analysis,'' \emph{IEEE Commun. Lett.}, vol. 28, no. 12, pp. 2930--2934, Dec.
2024.

\bibitem{survey_noisebased}
H. T. P. Da Silva \emph{et al.}, ``A survey on noise-based communication,''
\emph{IEEE Access}, vol. 14, pp. 14722--14734, 2026.

\bibitem{wong2021fas}
K.-K. Wong, A. Shojaeifard, K.-F. Tong, and Y. Zhang, ``Fluid antenna
systems,'' \emph{IEEE Trans. Wireless Commun.}, vol. 20, no. 3, pp.
1950--1962, Mar. 2021.

\bibitem{khammassi2023}
M. Khammassi, A. Kammoun, and M.-S. Alouini, ``A new analytical
approximation of the fluid antenna system channel,'' \emph{IEEE Trans.
Wireless Commun.}, vol. 22, no. 12, pp. 8843--8858, Dec. 2023.

\bibitem{new2024insights}
W. K. New, K.-K. Wong, H. Xu, K.-F. Tong, and C.-B. Chae, ``Fluid antenna
system: New insights on outage probability and diversity gain,'' \emph{IEEE
Trans. Wireless Commun.}, vol. 23, no. 1, pp. 128--140, Jan. 2024.

\bibitem{ramirez2024block}
P. Ram\'irez-Espinosa, D. Morales-Jimenez, and K.-K. Wong, ``A new spatial
block-correlation model for fluid antenna systems,'' \emph{IEEE Trans.
Wireless Commun.}, vol. 23, no. 11, pp. 15829--15843, Nov. 2024.

\bibitem{chai2022port}
Z. Chai, K.-K. Wong, K.-F. Tong, Y. Chen, and Y. Zhang, ``Port selection for
fluid antenna systems,'' \emph{IEEE Commun. Lett.}, vol. 26, no. 5, pp.
1180--1184, May 2022.

\bibitem{zhu2024fas_im}
J. Zhu, Q. Luo, G. Chen, P. Xiao, Y. Xiao, and K.-K. Wong, ``Fluid antenna
empowered index modulation for RIS-aided mmWave transmissions,'' \emph{IEEE
Trans. Wireless Commun.}, 2024.

\bibitem{stuber2018}
G. St\"uber, \emph{Principles of Mobile Communication}, 4th ed. Cham,
Switzerland: Springer, 2018.

\bibitem{zhang2002}
Q. T. Zhang and H. G. Lu, ``A general analytical approach to multi-branch
selection combining over various spatially correlated fading channels,''
\emph{IEEE Trans. Commun.}, vol. 50, no. 7, pp. 1066--1073, Jul. 2002.

\end{thebibliography}
\end{document}